\documentclass{acm_proc_article-sp}

\usepackage{url}
\usepackage{listings}
\usepackage{flushend}

\begin{document}

\newcommand{\whoknows}{\emph{WhoKnows?}}
\newcommand{\risq}{\emph{RISQ!}}
\newcommand{\patchr}{\emph{PatchR}}

\title{Collaboratively Patching Linked Data}
%
%
%
%
%

\numberofauthors{1} 
%
\author{
%
%
\alignauthor
Magnus Knuth, Johannes Hercher, Harald Sack\\
	\affaddr{Hasso-Plattner-Institute Potsdam}\\
	\affaddr{Prof.-Dr.-Helmert-Str. 2--3}\\
	\affaddr{14482 Potsdam, Germany}\\
	\email{ \{magnus.knuth$|$johannes.hercher$|$harald.sack\}@hpi.uni-potsdam.de}
}
\date{30 July 1999}

\maketitle

\begin{abstract}
Today's Web of Data is noisy. Linked Data often needs extensive preprocessing to enable efficient use of heterogeneous resources. While consistent and valid data provides the key to efficient data processing and aggregation we are facing two main challenges: (1st) Identification of erroneous facts and tracking their origins in dynamically connected datasets is a difficult task, and (2nd) efforts in the curation of deficient facts in Linked Data are exchanged rather rarely.
Since erroneous data is often duplicated and (re-)distributed by mashup applications it is not only the responsibility of a few original publishers to keep their data tidy, but progresses to become a mission for all distributers and consumers of Linked Data, too. We present a new approach to expose and to reuse patches on erroneous data to enhance and to add quality information to the Web of Data. The feasibility of our approach is demonstrated in the example of a collaborative game that patches statements in DBpedia data and provides notifications for relevant changes.
\end{abstract}

\category{H.3.5}{Information Storage and Retrieval}{On-line Information Services} [Data sharing]
\category{C.2.4}{Distributed Systems}{Distributed databases}
\category{I.2.4}{Computing Methodologies}{Knowledge Representation Formalisms and Methods}

\keywords{Linked Data, Crowdsourcing, Data Cleansing, Games With a Purpose, User Feedback Management, DBpedia} 

\section{Introduction}
\label{sec:introduction}

With the continuous growth of Linked Data on the World Wide Web (WWW) and the increase of web applications that consume Linked Data, the quality of Linked Data resources has become a relevant issue. Recent initiatives (as e.\,g., the Pedantic Web group\footnote{\url{http://pedantic-web.org/}}) uncovered various defects and flaws in Linked Data resources. Apart from structural defects, semantic flaws and factual mistakes are hard to detect by automatic procedures and require updates on the schema level, as well as on the data level.
Today, semantic web applications often rely on a local copy of originally distributed Linked Data due to system stability, access time, and data integration issues. When a genuine dataset is updated by the original publisher, changes made to a local copy may not be considered and get lost.

It is in fact a problem that erroneous data is distributed, duplicated, and reused in various semantic web applications, but it also opens up opportunities such as crowdsourcing to improve data quality. 
For example, a semantic web application might offer the possibility of user feedback to signalize facts, which need to be revised. Then, detected errors could be shared with the original data publisher or with other users of the dataset. Both would be able to update the identified defects.
While the need of error correction and data cleansing has reached the interest of the Linked Data community there exists no generally accepted method to expose, advertise, and retrieve suitable updates for Linked Data resources.
In order to reuse curation efforts and to realize the vision of a collaborative method for error detection and effective exchange of corresponding patches the following requirements have to be considered:
\begin{enumerate}
  \item The description of defects and their corresponding pat\-ches for Linked Data resources should be selectable by means of various criteria, as e.\,g., the scope of a patch, their advocates, provenance, and the type of defect to select patches efficiently.
  \item The realization of an appropriate workflow that covers guidelines to publish detected errors has to notify the original publishers as well as other users of a particular dataset. To encourage acceptance the execution of updates has to be as convenient as possible.
  \item Patches for Linked Data resources should also be published as Linked Data to ease their exchange and to make them available for rating, discussions, and reuse.
\end{enumerate}

The remainder of this paper is organized as follows. An overview of related work in the areas of Linked Data curation, crowdsourcing approaches, and related ontologies is provided in Section~\ref{sec:related}. In Section~\ref{sec:approach}, a new approach is discussed that allows to expose, rate, and select updates for particular Linked Data resources with a recommended ontology that is described in Section~\ref{sec:ontology}. The feasibility of this approach is illustrated exemplary in Section~\ref{sec:usecase}, where flaws detected with the help of a collaborative data cleansing game (\whoknows{} \cite{Waitelonis2011c}) are exposed and shared using the herein described ontology.
Section~\ref{sec:conclusion} concludes the paper with some general usage guidelines to encourage others to share curation efforts with the Linked Data community.

\section{Related Work}
\label{sec:related}

In order to raise quality in Linked Data applications various work has concentrated on syntactical and logical data consistency by providing validators for the Semantic Web languages RDF and OWL, e.\,g., W3C's RDF Validator\footnote{\url{http://www.w3.org/RDF/Validator/}}, Vapour\footnote{\url{http://validator.linkeddata.org/vapour}}, and OWL Validator\footnote{\url{http://owl.cs.manchester.ac.uk/validator/}}. Data quality in Linked Data has been criticized, as e.\,g., Hogan et al. analyzed typical errors and set up the RDF:Alerts Service\footnote{\url{http://swse.deri.org/RDFAlerts/}} that detects syntax and datatype errors to support publishers of Linked Data \cite{Hogan2010}. Nevertheless, recent analyses provided by LODStats show that there is still a vast amount of erroneous Linked Datasets available \cite{Demter2012}. However, validators are restricted to syntactical consistency and thus are not able to fulfill the following requirements: 

\begin{enumerate}
  \item Recognize the inconsistent usage of properties having restricted domains and ranges with entities belonging to another class. As for example, the RDF triple \\ \hfill \texttt{dbp:Apple\_Inc.~dbo:keyPerson dbp:Chairman .}\\
  is syntactically correct, but the range restrictions of \texttt{dbo:keyPerson} implies the entity \texttt{dbp:Chairman} to be type of class \texttt{dbo:Person}. While \texttt{dbp:Chairman} is untyped, it is rather a business role and from a user's point of view this might be incorrect because an actual person entity is expected to be a key person of a company.
  \item Recognize false facts that do not correspond to the reality, as e.\,g., the birthdate of a person can be syntactically totally correct, though factual wrong.
\end{enumerate}

Although validators are useful to verify syntactical consistency and correctness, they can not detect semantic or factual mistakes that may seem evident to a human. Therefore, an effective integration of human intelligence, i.\,e. crowdsourcing, is required. We address this issue by enabling interoperable exchange of user feedback on Linked Data facts. So far, this concept is sparsely present in Linked Data community.

A number of games with a purpose (GWAP) \cite{Ahn2004} are available that harness human intelligence for various complex tasks by providing appropriate incentives, as e.\,g., fun, competition, reputation, etc.~\cite{Siorpaes2010}. Most GWAP's are social web applications designed for the generation of metadata such as for multimedia content, but do not necessarily publish Linked Data.
Harnessing human intelligence for creating semantic content has been studied by Siorpaes and Simperl, who provide a collection of games\footnote{\url{http://ontogame.sti2.at/games/}} to build ontologies, annotating videoclips, or matching ontologies \cite{siorpaes2008,Siorpaes2010}. However, these games generally concentrate on content enrichment rather than on content curation. 

\whoknows{} is a simple quiz game in the style of `Who Wants to Be a Millionaire?' published previously by our research group, and designed to detect errors and shortcomings in DBpedia resources\cite{Waitelonis2011c}. Likewise, \risq{}\footnote{\url{http://apps.facebook.com/hpi-risq/}} is a game in the style of `Jeopardy!' that focusses on the evaluation and ranking of Linked Data properties about famous people. Both games are already well accepted but lack a standardized method to publish the obtained curation efforts.

We propose a Linked Data approach to describe changes made to Linked Data resources in order to make these updates (additions or deletions) also usable by the community. Therefore, we also have to consider work on data provenance information and version control on Linked Data resources. With respect to provenance, the \textit{Provenance Vocabulary}\cite{Hartig2009} (prv)\footnote{\url{http://trdf.sourceforge.net/provenance/ns.html}} is designed to make the origin, creation, and alteration of data transparent, but lacks concepts that describe the update of particular RDF triples. For the description of graph updates several ontologies have been defined, e.\,g., \textit{changeset} (cs)\footnote{\url{http://vocab.org/changeset/schema.html}}, the \textit{graph update ontology} (guo)\footnote{\url{http://webr3.org/specs/guo/}}, \textit{delta}\footnote{\url{http://www.w3.org/DesignIssues/Diff}}, and the \textit{triplify update vocabulary}\footnote{\url{http://triplify.org/vocabulary/update}}. All of these are limited to express changes within semantic data or to describe provenance of changes. 
Both approaches are not designed to promote, discuss, and exchange curation efforts, since means to express relevance as well as different types of updates do not exist. To our best knowledge none of these ontologies have been combined or extended to exchange curation information about Linked Data resources. 

In this paper a new approach is proposed to curate Linked Data collaboratively, as e.\,g., flaws in DBpedia resources that are hard to detect by automatic procedures. With respect to data cleansing of DBpedia resources one could argue that curation efforts should be applied directly to the original sources, i.\,e.~to the online encyclopedia Wikipedia\footnote{\url{http://www.wikipedia.org}}. 
The herein proposed approach goes beyond such efforts, as e.\,g., DBpedia Live \cite{Stadler2010} and can be applied to any Linked Data resource.
Furthermore, the method proposed in this paper supports the common practice to replicate original data resources at local repositories, and therefore enables a more convenient handling of data in regards of performance, stability, and integration issues.

\section{Workflow description}
\label{sec:approach}

By design, data providers and consumers at the Web of Data are not always the same party. Linked Data promotes to use external data resources within own applications. 
As a result, the datasets are not under control of the agent who employs the dataset, so that he could fix identified inconsistencies by himself. An alternative would be to request the dataset provider for a fix. 
Though, nowadays it is still common practice to set up an own data store containing web data as a local copy, may it be for reasons concerning performance or data control.

Either way, there is currently no standardized method to inform other parties using or distributing a particular data set about inconsistencies that have been detected within the data. We therefore suggest the \patchr{} vocabulary (cf. Section~\ref{sec:ontology}) that allows to describe patch requests including provenance information.

\begin{figure*}[hbt]
\centering
  \includegraphics[width=.65\textwidth]{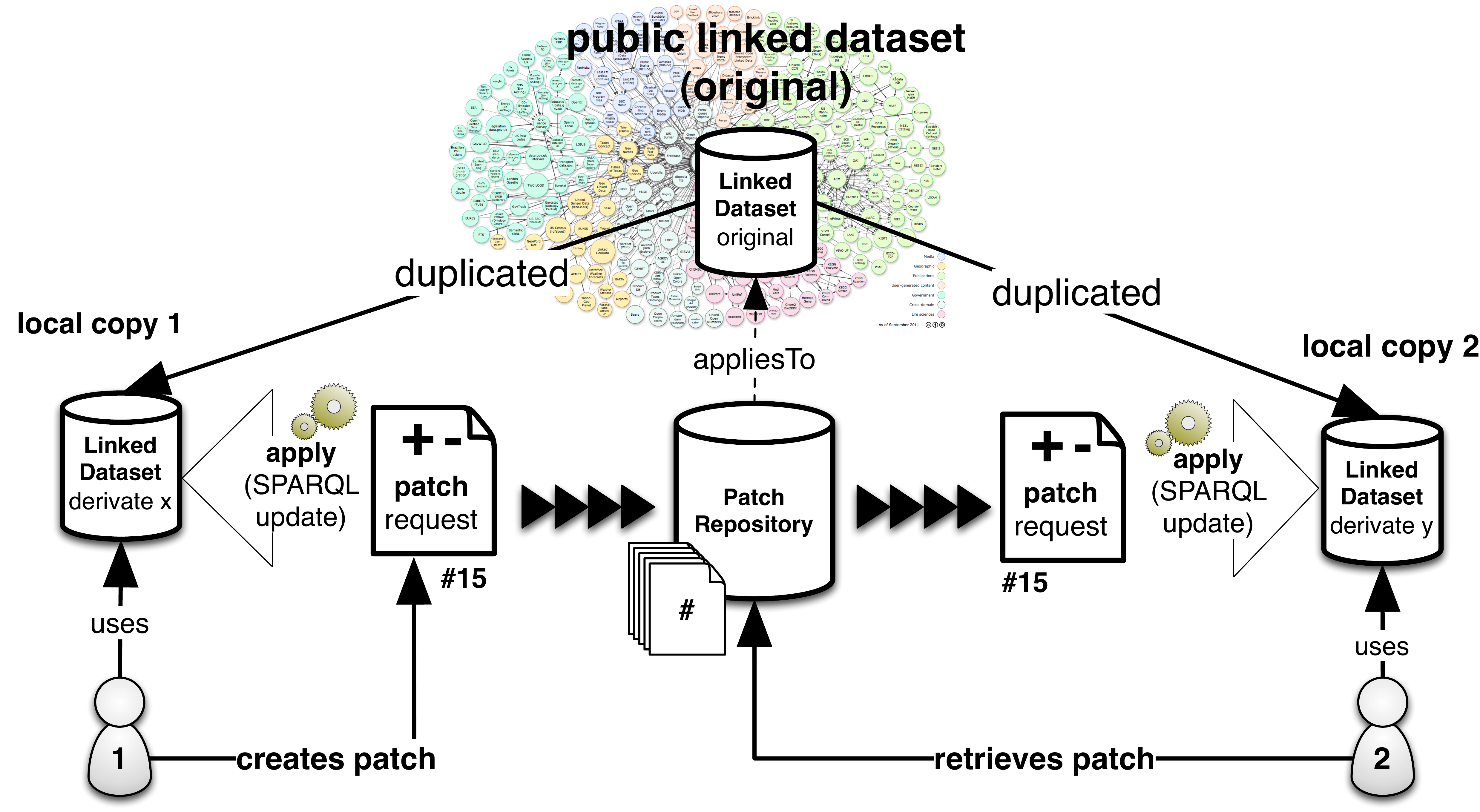}
  \caption{Workflow diagram for creating and applying a patch}
  \label{fig:workflow}
\end{figure*}

As shown in the workflow diagram in Fig.~\ref{fig:workflow}, multiple agents independently make use of a particular dataset, accessing it directly or as a local copy. 
Whenever an agent, whereby the agent may be human or an algorithm, identifies inconsistent facts (RDF triples) in the dataset, he can create a patch request.
The patch request describes the update that has to be performed on the dataset to solve the identified issue. As illustrated with the example in Listing~\ref{lst:patch} an update consists of a set of triples to add and/or a set of triples to delete in the dataset.

We suggest to publish patch requests at a 
patch repository. Though, everybody can easily set up their own repository, this would imply unnecessary administration effort. We propose a common centralized repository at least for each dataset, where multiple agents can report their findings to gain maximum feedback.
Since patch requests are encoded as Linked Data themselves, the repository represents an RDF graph, so that data about patch requests can be retrieved using the SPARQL query language.

Dataset providers, including providers of local copies can use the repository to retrieve individual updates for their hosted datasets. Dataset consumers can lookup patches to draw conclusions about the quality of a particular dataset. With simple modifications the repository can be extended to allow commenting and user voting for patches.

To extract an update for a particular dataset, the repository can be queried for patches having adequate quality constraints according to the required level of trust, as e.\,g. having a minimum number of supporters.
The resulting patches can directly be transformed into SPARQL update queries to enable a convenient update of the dataset.

\section{Description of the Patch Request Ontology}
\label{sec:ontology}

The Patch Request Ontology (pro)\footnote{cf. \url{http://purl.org/hpi/patchr}}, subsequently referred to as \patchr{}, provides a straightforward method to share user feedback about incorrect or missing RDF triples.
By wrapping the \texttt{guo:UpdateInstruction} concept adopted from the \emph{Graph Update Ontology}\footnote{cf. \url{http://webr3.org/owl/guo}} in a \texttt{pro:Patch} a \texttt{foaf:Agent} might publish requests to add, modify, or delete particular facts from a dataset.
Each patch is described by provenance information and a dataset to which it applies. 
Furthermore, the ontology supports a simple classification of patches using the \texttt{pro:patchType} property to allow convenient retrieval of common patches. These patch types may refer to commonly observed errors, as e.\,g., encoding problems or factual errors.

\begin{figure}[hbt]
 \begin{center}
  \includegraphics[width=.5\textwidth, trim=0 2cm 2cm 4cm, clip]{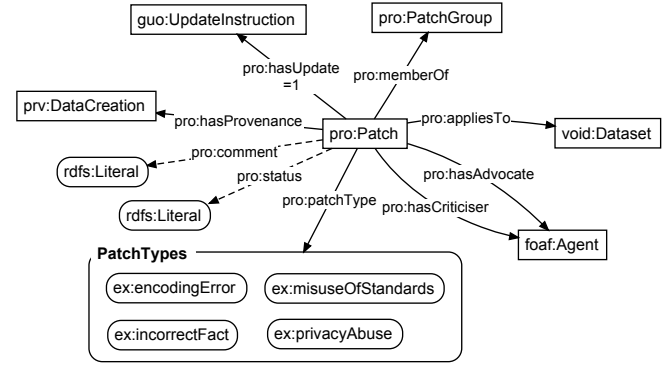}
  \caption{Overview on the patch request ontology}
  \label{fig:ontology}
 \end{center}
\end{figure}

Groups of patches can be created to bundle individual pat\-ches, as e.\,g., of a particular service that apply to common problems or have relevance only for specialized domains or regions. Fig.~\ref{fig:ontology} provides an overview on the main concepts of the \patchr{} ontology, which are described in Table~\ref{tab:ontology}.

\begin{table*}[ht]
\caption{Description of properties in the patch request ontology}
\centering
\begin{tabular}{p{.15\textwidth}|p{.7\textwidth}}
Property		& Description\\
\hline
hasUpdate		& Refers to a \texttt{guo:UpdateInstruction}. There can be only one UpdateInstruction per patch.\\
hasProvenance	& Refers to the provenance context, where this patch was created.\\
memberOf		& Assignment of a patch to a patch group.\\
appliesTo		& Refers to a \texttt{void:Dataset}, to allow convenient selection of patches.\\
hasAdvocate		& A link to people who support a submitted patch.\\
hasCriticiser	& A link to individual entities that disagree with the purpose of this patch.\\
patchType		& This property refers to a classification of a patch. A patch can have multiple types but should have at least one.\\
comment			& An informational comment on the patch.\\
status			& The status of a patch, as e.\,g., `active' or `resolved'. \\
\end{tabular}
\label{tab:ontology}
\end{table*}

\section{A Use Case for PatchR}
\label{sec:usecase}

As a use case for the \patchr{} ontology we extended \whoknows{} \cite{Waitelonis2011c}, an online quiz game in the style of `Who Wants to Be a Millionaire?' that generates questions from DBpedia facts\footnote{The currently deployed instance of \whoknows{} is based on DBpedia version 3.5.1}. The game can either be played via Facebook\footnote{\url{http://apps.facebook.com/whoknows_/}} or standalone\footnote{\url{http://tinyurl.com/whoknowsgame}}.
The game principle is to present multiple choice questions to the user that have been generated out of facts from DBpedia RDF triples. The player scores points by giving correct answers within a limited amount of time and loses lives when giving a wrong answer.

\begin{figure}[hbt]
\centering
\parbox{.5\textwidth}{
  \centering
  \includegraphics[width=.49\textwidth]{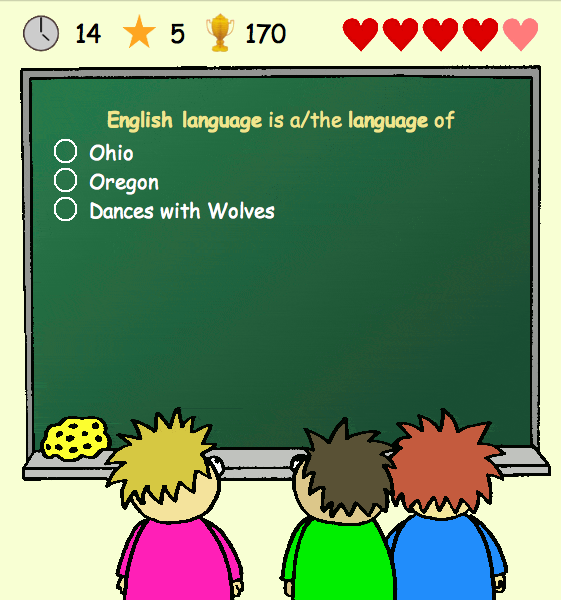}

\begin{tabular*}{.5\textwidth}{@{\extracolsep{\fill}}|c|c|c|}
	\hline
	{\footnotesize Subject} & {\footnotesize Property} & {\footnotesize Object}\\
	\hline \hline
	{\bf Ohio} & {\bf language} & {\bf English language}\\
	\hline
	{\footnotesize Oregon} & {\footnotesize language} & {\footnotesize De jure}\\
	\hline
	{\footnotesize Dances with Wolves} & {\footnotesize language} & {\footnotesize Lakota language}\\
	\hline
\end{tabular*}
}
\caption{Screenshot and triples used to generate a One-To-One question}
\label{fig:english}
\end{figure}

As an example, Fig.~\ref{fig:english} shows the question \emph{`English language is the language of ...?'} with the expected answer \emph{`Ohio'}. The question originates from the triple \medskip\\ \hfill
\medskip\makebox[.5\textwidth][c]{\texttt{dbp:Ohio dbo:language dbp:English\_language .}}
and is composed by turning the triple's order upside down: \textit{`Object is the property of: subject1, subject2, subject3...'}. The remaining choices are selected from subjects applying the same property at least once, but are not linked to the given object.
When the player submits her answer, the result panel will once again show all choices whereby the expected answer is highlighted. For each entity used in the question, a link to the respective DBpedia page is provided, whereby the user might examine the resource.

\begin{figure}[htb]
 \begin{center}
  \includegraphics[width=.49\textwidth]{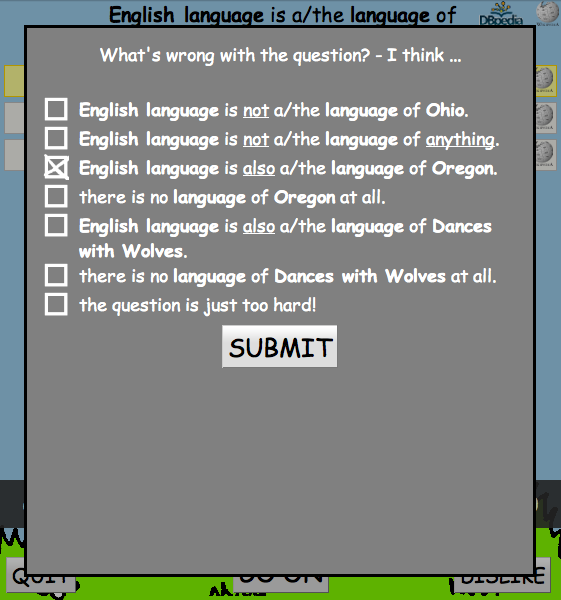}
  \caption{Screenshot of \whoknows{}' refinement panel.}
  \label{fig:refine}
 \end{center}
\end{figure}

So far, it was already possible for the players to simply report odd or obviously wrong questions by selecting a general `Dislike' button. We experienced that `disliked' questions often arose from inconsistency, i.\,e. wrong or missing triples. Therefore, this feature has been extended in a way that players can specify the particular fact, which they think to be incorrect by selecting it from a given set of potential inconsistencies. Those potential inconsistencies are presented to the user as natural language sentences. Fig.~\ref{fig:refine} shows a screenshot of this refinement panel. Sentences of the form \textit{`Object is \underline{not} a property of subject.'} indicate a wrong fact in the dataset, while sentences of the form \textit{`Object is \underline{also} a property of subject.'} indicate a missing fact.
From this user vote the system generates a patch request for either
\begin{itemize}
 \item deleting one or several triples or
 \item inserting one or several triples in the underlying knowledge base.
\end{itemize}

Listing~1 shows an example description of a patch request created by the game demanding the insertion of the triple \medskip \\ \hfill
\medskip\makebox[.5\textwidth][c]{\texttt{dbp:Oregon dbo:language dbp:English\_language .}}
The created patch requests are immediately posted to a patch repository implemented as a standard triplestore. The patch requests collected in the repository are publicly available and can be accessed via a SPARQL endpoint\footnote{\url{http://purl.org/hpi/patchr-sparql}}.

\lstset{
  basicstyle=\ttfamily\footnotesize,
  numbers=left,
  numberstyle=\tiny,
  numbersep=5pt,                  
  }
\begin{lstlisting}[caption={An example patch request},label={lst:patch}]
repo:Patch_15 a pro:Patch ;
  pro:hasUpdate [ 
    a guo:UpdateInstruction ;
    guo:target_graph <http://dbpedia.org/> ; 
    guo:target_subject dbp:Oregon ;
    guo:insert [ 
      dbo:language dbp:English_language ] 
    ] ;
  pro:hasAdvocate repo:Player_25 ;
  pro:appliesTo 
     <http://dbpedia.org/void.ttl#DBpedia> ;
  pro:status "active" ;
  pro:hasProvenance [ 
    a prv:DataCreation ;
    prv:performedBy repo:WhoKnows ;
    prv:involvedActor repo:Player_25 ;
    prv:performedAt "..."^^xsd:dateTime ] .
\end{lstlisting}

To illustrate the collection of patch requests submitted to the repository, we implemented a user interface\footnote{\url{http://purl.org/hpi/patchrui}} that creates reports about the most recent and most popular patch requests.
Furthermore, inconsistencies for single entities of the DBpedia dataset can be displayed.

Patching DBpedia can be regarded as a special case, since this data is based on editable Wikipedia pages. Identified problems should sustainably be fixed directly in Wikipedia, so they won't occur in future DBpedia releases. Therefore, we also provide a link to the particular Wikipedia page, where users might solve the inconsistency directly by editing the relevant section directly.

Data providers interested in patches for the DBpedia dataset can query the patch collections for those patches they think are evident enough to apply to their local graph. To apply those patches directly onto a local triplestore, a set of SPARQL update queries can be generated. Listing~2 shows the update query corresponding to the preceding example.

\begin{lstlisting}[title={Listing~2: The appropriate SPARQL update query for Listing~1},label={lst:sparqlupdate}]
INSERT DATA INTO <http://dbpedia.org/> {
  dbp:Oregon
     dbo:language dbp:English_language .
}
\end{lstlisting}




\section{Conclusion and Outlook}
\label{sec:conclusion}

In this paper a collaborative approach is demonstrated that can leverage the quality in Linked Data applications by combining crowdsourcing methods with Linked Data principles. We have illustrated the feasability of this approach by a use case on top of a quiz game that collects data about possible flaws in DBpedia and publishes patches using a patch request ontology. Finally, a patch repository is presented allowing a convenient selection of appropriate updates on Linked Data resources.

As in our use case, some datasets are based on genuine information sources that are worth to be corrected directly at their origin to avoid the reoccurrence of the same errors in future releases. We also encourage such efforts for Wikipedia. Since Wikipedia pages must be edited by humans the patch repository's user interface supports web users by providing direct links to the respective DBpedia and Wikipedia pages that contain the facts to be improved.

However, our approach is not limited to the given showcase, but can be applied to any dataset that obeys Linked Data principles. Moreover, the approach does not rely on sophisticated crowdsourcing methods, but can also be adopted by algorithmic data curation systems.
Therefore, the presented application can be of concern for various data providers that are interested in data curation issues.
In general each aggregator of Linked Data can help to leverage structural and factual quality of the semantic web. The following steps are necessary for active participation:

\begin{enumerate}
 \item Identify potential flaws in original or aggregated data.
 \item Implement facilities to gather user feedback.
 \item Serialize identified flaws and corresponding updates using the \patchr{} ontology.
 \item Publish these patches within an appropriate repository that can be publicly accessed.
\end{enumerate}

In regards to managing distributed information on patches we suggest a rather centralized setting, where major dataset providers rely on dedicated patch repositories to obtain pat\-ches for their particular dataset. Further auxiliary tasks for effective synchronization of patches such as further standardization and management of trust are not covered in this publication and subject of future research.

The repository currently represents merely a proof of concept and various extensions are conceivable. Further work will include the implementation of advanced trust and access control mechanisms as well as validity checking. Features like rating, feedback, and reputation management are necessary to provide appropriate incentives in the long run. A pingback mechanism might be valuable to inform data providers about recently created patch requests concerning their datasets. Finally, also the implementation of an API can be considered to ease the publication of patches.

\bibliographystyle{splncs}
\bibliography{bibliography}

\begin{thebibliography}{1}

\bibitem{Waitelonis2011c}
Waitelonis, J., Ludwig, N., Knuth, M., Sack, H.:
\newblock {WhoKnows? - Evaluating Linked Data Heuristics with a Quiz that
  Cleans Up DBpedia}.
\newblock International Journal of Interactive Technology and Smart Education
  (ITSE) \textbf{8}(3) (2011)

\bibitem{Hogan2010}
Hogan, A., Harth, A., Passant, A., Decker, S., Polleres, A.:
\newblock {Weaving the Pedantic Web}.
\newblock In: Proc. of the Linked Data on the Web {(WWW2010)} Workshop {(LDOW}
  2010), Raleigh, North Carolina, {USA} (April 2010)

\bibitem{Demter2012}
Demter, J., Auer, S., Martin, M., Lehmann, J.:
\newblock {LODStats -- An Extensible Framework for High-performance Dataset
  Analytics}.
\newblock unpublished, available at: http://aksw.org/projects/LODStats
  (February 2012)

\bibitem{Ahn2004}
von Ahn, L., Dabbish, L.:
\newblock Labeling images with a computer game.
\newblock In: CHI '04: Proceedings of the SIGCHI conference on Human factors in
  computing systems, New York, NY, USA, ACM (2004)  319--326

\bibitem{Siorpaes2010}
Siorpaes, K., Simperl, E.:
\newblock Human intelligence in the process of semantic content creation.
\newblock World Wide Web \textbf{13} (2010)  33--59

\bibitem{siorpaes2008}
Siorpaes, K., Hepp, M.:
\newblock Games with a purpose for the semantic web.
\newblock IEEE Intelligent Systems \textbf{23}(11) (May 2008)  50--60

\bibitem{Hartig2009}
Hartig, O.:
\newblock Provenance information in the web of data.
\newblock In: LDOW2009, April 20, 2009, Madrid, Spain. (April 2009)

\bibitem{Stadler2010}
Stadler, C., Lehmann, J., Hellmann, S.:
\newblock {Update Strategies for DBpedia Live}.
\newblock Volume 699 of CEUR Workshop Proceedings. (February 2010)

\end{thebibliography}

\balancecolumns
\end{document}